\documentclass[a4paper, fontsize=12pt]{article}

\renewenvironment{abstract}
               {\list{}{\rightmargin\leftmargin}%
                \item[\hspace*{1cm}\small\textbf{Abstract ---}]\relax}
               {\endlist}

\usepackage[top=0.8in, bottom=0.8in, left=0.8in, right=0.8in]{geometry}
\usepackage[fleqn]{amsmath}
\usepackage{amssymb}
\usepackage{enumerate}
\usepackage{amsthm}
\usepackage[pdftex]{graphicx}
\usepackage{balance}
\usepackage{fancyhdr}
\usepackage{sidecap}
\sidecaptionvpos{figure}{t}
\usepackage{appendix}
\usepackage{float}
\usepackage{scrextend}
\usepackage{changepage}
\usepackage{endnotes}
\usepackage{url}
\usepackage{setspace}
\usepackage{hyperref}

\DeclareSymbolFont{symbolsC}{U}{txsyc}{m}{n}
\DeclareMathSymbol{\strictif}{\mathrel}{symbolsC}{74}

\let\footnote=\endnote

\newtheorem{Theorem}{Theorem}

\newtheorem{Reply}[Theorem]{Reply}

\begin{document}

\title{\bf A category mistake in observational claims\\ regarding ultrashort-lived unstable particles}

\date{}

\author{\small{Marcoen J.T.F. Cabbolet}\\
        \small{\textit{Department of Philosophy, Free University of Brussels}}\\
        \small E-mail: Marcoen.Cabbolet@vub.ac.be
        }
\maketitle

\begin{abstract}\small
The physics literature contains many claims that ultrashort-lived unstable particles, such as a Higgs boson, have been observed. These claims are a matter of applying the $5\sigma$-convention in particle physics. This paper, however, shows that by applying this $5\sigma$-convention a category mistake is made, by which a \emph{pure reasoning} is passed off as an \emph{observation}. Not only are these two fundamentally different primitive notions at the very basis of science, but the pure reasoning in question is also weaker than an observation: what we have in each case is that the existence of the ultrashort-lived unstable particle is inferred to the best explanation, but that does \emph{absolutely not} merit the stronger claim that the particle in question has been ``observed''. Consequently, the observational claims in question will thus have to be dismissed as overstatements. On a general note, this demonstrates that the empirical support for the Standard Model of particle physics is significantly less than hitherto thought.
\end{abstract}

\section{Introduction}
\begin{SCfigure}[1.0][h]
\centering
\includegraphics[width=0.5\textwidth]{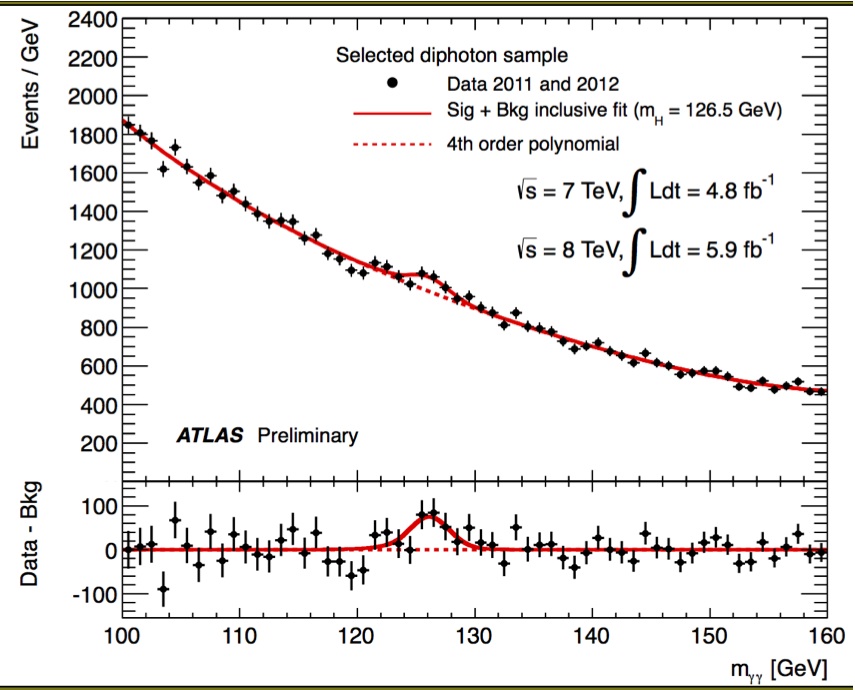}
    \caption{
   Diphoton mass spectrum obtained in the hunt for the Higgs boson. The lower curve with the peak at around 125 GeV is obtained from the upper one by substraction. Source: CERN Document Server.
  }
\label{fig:1}
\end{SCfigure}

\noindent In recent decades, claims have surfaced in the physics literature that ultrashort-lived unstable particles, postulated by the Standard Model c.q. in the framework of the Standard Model, have been ``observed''. Examples are the claims regarding the Higgs boson \cite{ATLAS,CMS,CERN}, the $W^{\pm}$ bosons \cite{CERN1983a}, the $Z^0$ boson \cite{CERN1983b,CERN1983c}, the Y-meson \cite{E288}, the J/$\Psi$ meson \cite{Aubert}, the $\Omega^-_b$ baryon \cite{DO}, and the tetraquark $Z(4430)^-$ \cite{LHCb}. It is then important to realize that the use of the word `observation' (sometimes replaced by `discovery', meaning `first-time observation') is the key aspect of these papers: it is \emph{therefore} that the existence of these particles is widely believed to be confirmed. The method by which physicists reach the conclusion that the sought-after ultrashort-lived unstable particle has been ``observed'' is the same in all cases. Experimentally, each decay mode of the ultrashort-lived unstable particle is analyzed separately: by each such analysis one tests the hypothesis `predicted-decay-products-exist' versus the hypothesis `no-predicted-decay-product'. E.g. in the hunt for the Higgs boson, the diphoton mass spectrum shown in Fig. \ref{fig:1} was obtained in the analysis of the decay mode $H\rightarrow\gamma\gamma$: one then accepts the hypothesis `the 125 GeV photon pairs predicted by Higgs decay exist', and rejects the hypothesis `the predicted 125 GeV photon pairs do not exist'.
If all goes well, the conjunction of accepted hypotheses obtained from the analyses yields the \emph{intermediate conclusion} that the predicted decay products of the ultrashort-lived unstable particle have been observed with a significance of $5\sigma$. None of this is questioned: the calculations involved in deriving testable predictions, the experimental work itself and the statistical analyses of the experimentally obtained data are all state-of-the-art. But the problematic step comes thereafter: from this intermediate conclusion physicists reach the \emph{final conclusion} that the ultrashort-lived unstable particle \emph{itself} has been observed by applying the following convention in particle physics, which will henceforth be called the `$5\sigma$-convention':
\begin{quote}
{\it the observation of an ultrashort-lived unstable particle can be claimed if the condition is satisfied that its predicted decay products have been observed with a significance of $5\sigma$}.
\end{quote}
The purpose of this paper is to show that when this $5\sigma$-convention is applied, a \emph{pure reasoning} is passed off as an \emph{observation}---two primitive notions at the very basis of science that since long have been established to be fundamentally different, e.g. by the now historical works of Bacon, Descartes, Hobbes, Hume, Locke and the likes thereof. So, applying this $5\sigma$-convention yields a wrong use of primitive notions of science---it yields a \emph{category mistake}, which has nothing to do with measurement or calculation, but which \emph{overstates} results.
\section{Methodology}
Physicists at CERN have been confronted with the criticism in \cite{Cabbolet} that the $5\sigma$-convention doesn't resonate with recent philosophical insights in what it means to have observed something \cite{VanFraassen,Fox}, but the reply was that the physicists' use of the term `observation' should not be confused with potentially different definitions of `observation' in philosophy (D. Schlatter, CERN, personal communication). So, to prove that what physicists call an ``observation'' of an ultrashort-lived unstable particle is not an observation at all, below it will be proved that the condition laid down in the $5\sigma$-convention is \emph{insufficient} for an observational claim regarding an ultrashort-lived unstable particle. That can be done by a standard method. The logical form of the $5\sigma$-convention is, namely, that of an implication
\begin{equation}\label{eq:5sigma}
S \Rightarrow C
\end{equation}
where the proposition letter `$C$' stands for the desired conclusion that the observation of the ultrashort-lived unstable particle $X$ can be claimed, and `$S$' for the (allegedly) sufficient condition for that claim, being that the predicted decay products have been observed with a significance of $5\sigma$. To prove that the condition $S$ is \emph{insufficient}, it suffices to prove implications
\begin{gather}\label{eq:N}
C \Rightarrow N \\
\label{eq:negN}
S \Rightarrow \neg N
\end{gather}
for some proposition $N$: Eq. (\ref{eq:N}) means that $N$ is a necessary condition for $C$, and Eq. (\ref{eq:negN}) means that this necessary condition $N$ is not satisfied when $S$ is satisfied---that proves that the allegedly sufficient condition $S$ is \emph{insufficient}, and thus that the $5\sigma$-convention is \emph{inadequate}.

\section{Result: the argument against the $\mathbf{5\sigma}$-convention}
We first identify a necessary condition for an observational claim. The point is then this: regardless of how we define the term `observation' precisely, the crux is that it is an \emph{act of the senses}. Consequently, the following implication is true:
\begin{quote}
{\it \textbf{if} it can be claimed that $X$ has been observed, \textbf{then} it is necessarily true that $X$ exists---that is, \textbf{then} the real world can only be a world in which $X$ exists, and not one in which $X$ does not exist}
\end{quote}
This implication will henceforth be called `the major': Eq. (\ref{eq:N}) expresses its logical form, with the proposition letter `$N$' standing for `it is necessarily true that $X$ exists'---this is, thus, a necessary condition for any observational claim. (This major also holds for observations of other things than ultrashort-lived unstable particles.)

It can be proven by \emph{reductio ad absurdum} that this indeed is a necessary condition. Suppose that we deny the major: that would be admitting to the possibility that we can claim that $X$ has been observed, while $X$ does not exist. That is patently absurd. In a similar vein, Kant argued that we must realize that if we perceive a phenomenon, there must be a thing in itself whose appearance we perceive; ``[f]or, otherwise, we should require to affirm the existence of an appearance, without something that appears---which would be absurd'' (preface to the $2^{\rm nd}$ edition of \emph{Critique of Pure reason}, 1787). That proves that the denial of the above major leads to an absurdity: the major can, thus, \textbf{not} be denied---those who nevertheless deny the major are cordially invited on an expedition to spot a unicorn: if existence of a thing is not needed for its observation, we might spot one.\\
\ \\
Proceeding, we now prove that the necessary condition for an observational claim identified above is not satisfied when the allegedly sufficient claim of the $5\sigma$-convention is satisfied. In other words, we now prove that the following implication, which will henceforth be called `the minor', is true:
\begin{quote}
{\it \textbf{if} the predicted decay products of the ultrashort-lived unstable particle $X$ have been observed with a significance of $5\sigma$, \textbf{then} it is not necessarily true that $X$ exists---that is, \textbf{then} the real world can be a world in which $X$ exists, but also one in which $X$ does not exist}
\end{quote}
Eq. (\ref{eq:negN}) thus expresses the logical form of this minor. So, suppose that we have observed what can be described as the predicted decay products of the ultrashort-lived unstable particle $X$, with a significance of $5\sigma$. We then have to admit the simple truths (i) that $X$ itself has never been the sensum, i.e. the thing being sensed, in an \emph{act of the senses}, and (ii) that the existence of $X$ is not \emph{logically} implied by the empirical data. So, we are left with inference to the best explanation (IBE), which is an \emph{act of pure reasoning}: the \emph{only} conclusion that we can draw is that the existence of the ultrashort-lived unstable particle $X$ is the best explanation for the empirical data. But that means that it is not necessarily true that the ultrashort-lived unstable particle $X$ exists: the empirical data by themselves, namely, admit \emph{both} a real world in which the observed particles exist and $X$ exists, \emph{and} a real world in which the observed particles exist but $X$ does not exist. We may \emph{reason} that the first possibility is \emph{currently} the best explanation of the empirical data, but---regardless of how much we \emph{want} this to be true---by no means does this \emph{exclude} the second possibility. To illustrate this with an example, consider the case of the Higgs boson. This has the essential property $P$ that it `gives mass' to other particles. But this property $P$ is not reflected in its decay products: from merely observing the decay products, it cannot be logically concluded that the source is an ultrashort-lived unstable particle $X$ \textbf{with} the property $P$---the decay products may also have originated from a particle $X^\prime$ that has the same decay reactions as the Higgs boson but \textbf{not} the property $P$. So, \emph{even though} the existence of the Higgs boson is the best explanation now, the empirical data by themselves still admit the possibility that it does not exist. Concluding, even though the allegedly sufficient claim of the $5\sigma$-convention is satisfied, we are forced to admit that it is not necessarily true that the ultrashort-lived unstable particle $X$ exists. That proves the minor---it is emphasized that this minor obtains from an analysis of the mere concept of IBE.

It is important to realize that we engage in circular reasoning if we deny the minor---to deny the minor is to hold the view, expressed by the conjunction $S \wedge N$, that an observation of the predicted decay products of the ultrashort-lived unstable particle $X$ goes hand in hand with the presence of the ultrashort-lived unstable particle $X$ in the system under observation. What we have is that the analysis of the experimental data reveals that (an excess of) certain particles with certain properties must have been present in the system under observation, but the crux is that the analysis in itself doesn't reveal anything else: we may describe the observed particles as the predicted decay products of the ultrashort-lived unstable particle $X$, but it remains to be proven that the existence of the ultrashort-lived unstable particle $X$ is the \emph{cause} of their presence. So, if we take the position expressed by the conjunction $S \wedge N$, then we have tacitly assumed what has to be proven: we have, then, engaged in circular reasoning. So, the minor cannot be denied. This circular reasoning also obtains when we call it an \emph{indirect observation} of $X$: knowledge of the cause of the observed phenomenon is then, namely, presupposed \cite{Fox}. (The circular reasoning can be laid bare in a suitable frame for modal propositional logic, but that's beyond the scope of this paper.)\\
\ \\
Using standard propositional logic, the inevitable conclusion that can be drawn from the major and the minor is then the following implication:
\begin{quote}
{\it \textbf{if} the predicted decay products of the ultrashort-lived unstable particle $X$ have been observed with a significance of $5\sigma$, \textbf{then} it cannot be claimed that $X$ has been observed}
\end{quote}
The logical form of this conclusion is then expressed by the implication
\begin{equation}
S \Rightarrow \neg C
\end{equation}
in which the proposition letters `$S$' and `$C$' have the same meaning as above; it follows straight from Eqs. (\ref{eq:N}) and (\ref{eq:negN}). That establishes as a fact that the $5\sigma$-convention is \emph{inadequate}---the allegedly sufficient condition $S$ is \emph{insufficient} for an observational claim. So, by applying the $5\sigma$-convention a category mistake is made: what has actually been done is that the existence of $X$ has been inferred on the basis of IBE, which is an \emph{act of pure reasoning}, but this is (wrongly) passed off as an observation of $X$---that is, as an inference of the existence of $X$ on the basis of an \emph{act of the senses}. The crux is that $X$ has never been the `sensum' in an act of the senses!

\section{Discussion: replies by physicists and philosophers}

Below three replies---the first by a top philosopher and the last two by top physicists---to the above argument against the $5\sigma$-convention are discussed. The replies are paraphrased.

\begin{Reply}
The paper is correct. Yes, high energy physicists use the term `observation' in a misleading way in this context. But philosophically, the argument is not interesting.
\end{Reply}
\noindent This assessment is more or less correct: the purpose of this paper is merely to show that the use of the word `observation' by physicists is wrong---at least, in so far as it concerns claims that ultrashort-lived unstable particles have been ``observed''. The purpose of this paper is thus \textbf{not} to advance a more general philosophical discussion: consequently, this paper will be not interesting for those interested in more general philosophical discussions.

\begin{Reply}
This is not worthy of further consideration by physicists, because it is a mere epistemological treatise on the cognizability of short-lived particles such as the Higgs boson.
\end{Reply}

\noindent The assessment, that this is a mere epistemological treatise on the cognizability of short-lived particles, is also more or less correct. That, however, does not make the argument any less true. The point is that \emph{physicists} are the ones who have used the term `observation' wrongfully, and therefore the critical discussion about that word use has to be held \emph{with physicists}. Furthermore, since existential knowledge in physics, that is, knowledge that this or that exists, derives from observations, the obvious risk of this wrong use of terminology is the creation of a body of pseudoknowledge---i.e. a body of statements that is falsely believed to be a body of knowledge. Therefore, there is no point in postponing a critical discussion any further.

\begin{Reply}
I don't have the impression that this teaches physicists anything new. The author is correct when saying that an inference to the best explanation is involved; but everyone knows that. But if we are confident that the Higgs boson is the actual cause of the observed decay products, then we can be confident that we have indirectly observed the Higgs boson.
\end{Reply}

\begin{SCfigure}[1.0][h]
 \centering
\includegraphics[width=0.55\textwidth]{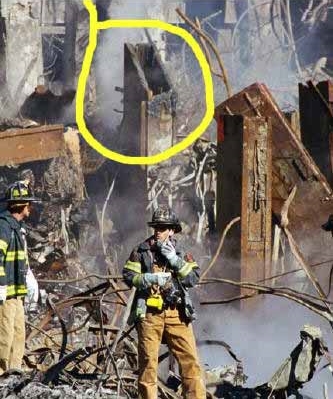}
    \caption{
   Photograph taken after collapse of the WTC towers. Enclosed in the yellow circle is a column with a straight cut. Source: public domain.
  }
\label{fig:3}
\end{SCfigure}

\noindent This physicist has missed the whole point of the presented argument: if you have merely inferred that the ultrashort-lived unstable particle $X$ exists because it is the best explanation for the observations, then you cannot pass that off as an observation of the ultrashort-lived unstable particle $X$ \emph{regardless of the degree of confidence that you have that these particles are the cause of the observed phenomenon}. By the same token, it cannot be claimed that on the photo in Fig. \ref{fig:3} below we ``observe'' government complicity in the 9/11 atrocities, \emph{regardless of one's degree of confidence that government complicity is the actual cause of the straight cut of the column}. In both cases, a reasoning is passed off as an observation. So only by lowering the standard of quality required for contributions to the scientific discourse we can allow the claim that a Higgs boson has been ``observed''---but then we will also have to allow the claim that government complicity in the 9/11 atrocities has been ``observed''!

\section{Conclusions}
First of all, this paper has proven that that $5\sigma$-convention, on which claims that ultrashort-lived unstable particles have been ``observed'' are based, is inadequate. The validity of this conclusion depends on nothing else than on the validity of two premises: the only way to prove that the present paper is incorrect is, thus, to prove that one of its premises is incorrect. But clear reasons have been given why the premises cannot be denied.

Furthermore, this is not a word game: the inadequateness of the $5\sigma$-convention has far reaching implications, not the least of which is that all observational claims based thereon have to be dismissed as \emph{overstatements}. Inferring the existence of an ultrashort-lived unstable particle $X$ on the basis of IBE is, namely, \emph{much weaker} than inferring the existence of $X$ on the basis of an act of the senses: in the latter case it necessarily true that $X$ exists in the system under observation, but in the former case it isn't. And \emph{because of} this non-equivalence, none of the ultrashort-lived unstable particles postulated by the Standard Model can be claimed to have been observed---at best it can be claimed in each case that the predictions of the Standard Model (including the ultrashort-lived unstable particle $X$) have been confirmed, which is a substantially different claim. Ergo, the empirical support for the Standard Model of particle physics is \emph{significantly less} than currently thought!

The present argument is strictly limited to observational claims concerning ultrashort-lived unstable particles: by no means is this intended to be applied to observational claims concerning things, living or lifeless, that can be directly observed---e.g. `cows exist'. However, the physics literature contains many more observational claims that in fact are category mistakes in which the \emph{observation of a thing} and the \emph{inference of the existence of a thing based on IBE} have been confused; a recent example is the claimed observation of a gravitational wave \cite{LIGO}, which led to the award of the 2017 Nobel prize in physics. The recommendation is therefore to reassess observational claims in the physics literature.

\end{document}